\documentclass[12pt,a4paper]{article}

\usepackage[usenames]{color}
\usepackage{hyperref}
\usepackage{epsfig,amsmath,wrapfig,amsthm,dsfont,bm,bbm}
\usepackage{mathrsfs}
\usepackage{amsfonts}
\usepackage{authblk}

\usepackage{graphicx}
\usepackage{caption}
\usepackage{subcaption}
\usepackage{natbib}
\usepackage{bm}

\usepackage{booktabs}

\usepackage{lineno}

\usepackage{tikz}
\usetikzlibrary{shapes.geometric, arrows}

\tikzstyle{io} = [rectangle, rounded corners, text centered, draw=black, fill=red!30]

\tikzstyle{startstop} = [trapezium, trapezium left angle=80, trapezium right angle=100, text centered, draw=black, fill=blue!30]

\tikzstyle{process} = [rectangle, text centered, draw=black, fill=orange!30]

\tikzstyle{decision} = [diamond, minimum width=3cm, minimum height=1cm, text centered, draw=black, fill=green!30]

\tikzstyle{arrow} = [thick,->,>=stealth]

\usepackage{graphicx}

\usepackage[paperwidth=8.5in, paperheight=11in, margin=1in]{geometry}
\usepackage{setspace}
\def\boldfacefake#1{\kern-4pt
   \hbox{ \mathsurround=0pt
   \hbox to 0.4pt{$#1$\hss}\hbox to 0.4pt{$#1$\hss}\hbox {$#1$}}}

\newcommand{\Var}{\mbox{Var}}

\newcommand{\btable}{\begin{table}[h]\centering}
\newcommand{\etable}{\end{table}}
\newcommand{\bt}{\begin{parag}\small \let\b=\nsb \let\sb=\nssb \begin{tabular}}
\newcommand{\et}{\end{tabular}\let\b=\nb \let\sb=\nsb\end{parag}}

\newenvironment{parag}{\par}{\par}

\newcommand{\be}{\begin{eqnarray}}
\newcommand{\ee}{\end{eqnarray}}
\newcommand{\ba}{\begin{eqnarray*}}
\newcommand{\ea}{\end{eqnarray*}}

\newtheorem{theorem0}{Theorem}
\newtheorem{lemma0}{Lemma}
\newtheorem{remark0}{Remark}
\newtheorem{fact0}{Fact}
\newtheorem{example0}{Example}
\newtheorem{definition0}{Definition}
\newtheorem{corollary0}{Corollary}
\newtheorem{proposition0}{Proposition}
\newtheorem{algorithmY}{Algorithm}

\newcommand{\reals}{\mbox{\rm I\kern-.20em R}}
\newcommand{\sreals}{\mbox{\small \rm I\kern-.20em R}}

\DeclareFontFamily{OT1}{pzc}{}
\DeclareFontShape{OT1}{pzc}{m}{it}{<-> s * [1.10] pzcmi7t}{}
\DeclareMathAlphabet{\mathpzc}{OT1}{pzc}{m}{it}

\makeatletter
\renewcommand*{\@fnsymbol}[1]{\ensuremath{\ifcase#1\or *\or *\or *\or
   \mathsection\or \mathparagraph\or \|\or **\or \dagger\dagger
   \or \ddagger\ddagger \else\@ctrerr\fi}}
\makeatother

\makeatletter
\if@compatibility
\renewenvironment{titlepage}
    {%
      \if@twocolumn
        \@restonecoltrue\onecolumn
      \else
        \@restonecolfalse\newpage
      \fi
      \thispagestyle{empty}%
    }%
    {\if@restonecol\twocolumn \else \newpage \fi
    }
\else
\renewenvironment{titlepage}
    {%
      \if@twocolumn
        \@restonecoltrue\onecolumn
      \else
        \@restonecolfalse\newpage
      \fi
      \thispagestyle{empty}%
      \setcounter{page}\@ne
    }%
    {\if@restonecol\twocolumn \else \newpage \fi
     \if@twoside\else
     \fi
    }
\fi
\makeatother

\title{
Multiple-trait Adaptive Fisher's Method for Genome-wide Association Studies 
}

\author[1]{Qiaolan Deng}
\author[1]{Chi Song}
\affil[1]{College of Public Health, Division of Biostatistics, The Ohio State University}

\begin{document}
\begin{titlepage}
\maketitle
\end{titlepage}

\section*{Abstract}
In genome-wide association studies (GWASs), there is an increasing need for detecting the associations between a genetic variant and multiple correlated traits, which are often measured together in a single GWAS. Despite the multivariate nature of the studies, single-trait-based methods remain the most widely-adopted procedure, owing to their simplicity. However, the association between a genetic variant and single traits sometimes can be weak to detect, and ignoring the actual correlation among traits may lose power. On the contrary, multiple-trait methods that analyze all the traits simultaneously are much more powerful. Although multiple-trait methods have been developed, several drawbacks limit their wide application. First, many existing methods can only process continuous traits and fail to allow for binary traits which are ubiquitous in real-world problems. Second, as shown in our simulation, the performance of many existing methods is unstable under different scenarios where the correlation among traits and the signal proportion vary. In this paper, we propose a multiple-trait adaptive Fisher's (MTAF) method to test associations between a genetic variant and multiple traits, by adaptively aggregating evidence from each trait. The proposed method can accommodate both continuous and binary traits and it has reliable performance under various scenarios. Using a simulation study, we compared our proposed method with several existing methods and demonstrated its competitiveness in terms of type I error and statistical power. We also applied the method to the Study of Addiction: Genetics and Environment (SAGE), and successfully identified several genes associated with substance dependence.





\section*{Introduction}
\label{sec:introduction}

Genome-wide association studies (GWASs) explore the associations between genetic variants, called single-nucleotide polymorphisms (SNPs), and traits \citep{visscher2012five}. They have been successfully applied to identify numerous genetic variants associated with complex human diseases \citep{buniello2019nhgri}. In GWASs, it is common to measure multiple traits underlying complex diseases, because due to pleiotropy one genetic variant can influence multiple phenotypic traits \citep{solovieff2013pleiotropy}. For example, many genetic variants are associated with both fasting glucose and fasting insulin of type 2 diabetes \citep{Billings2010}; GWASs found a variant in gene SLC39A8 that influences the risk of schizophrenia and Parkinson disease.\citep{pickrell2016detection}. In the last decade, single-trait methods have been widely adopted \citep{visscher201710} in which the association between the genetic variant and each single trait is tested one at a time. However, this type of method suffers from several disadvantages. First, sometimes the association between a single SNP and a trait is too weak to be detected by itself. Second, it ignores the correlation structure among the traits, which leads to the loss of statistical power when the traits are truly correlated. Third, the post-hoc combination of multiple tests without proper adjustment may lead to inflated type I error or compromised statistical power. As a result, there is an increasing need to develop powerful statistical methods that are capable of testing multiple traits simultaneously and properly.

Various statistical methods have been developed and applied to multiple-trait studies. Following an overview of multiple-trait methods by \citet{Yang2012}, we classify the existing methods into three categories. The first category is to combine test statistics or p-values from univariate tests. The O'Brien method \citep{o1984procedures, wei1985combining} combines the test statistics from the individual test on each trait weighted by inverse variance. The sum of powered score tests (SPU) \citep{Pan2014} and adaptive SPU (aSPU) \citep{Zhang2014} combines the score test statistics derived from generalized estimation equations (GEE). The Trait-based Association Test that uses Extended Simes procedure (TATES) \citep{van2013tates} exploits the correlation among p-values from each univariate test and generate a new test statistic. Fisher's method \citep{fisher1925statistical, yang2016efficient} and Cauchy's method \citep{liu2020cauchy} combines p-values of single-trait analyses and get the final p-values from known probability distributions. The second category is to reduce the dimensions of multiple traits. Principal components of heritability (PCH) \citep{Klei2008} collapses the multiple traits to a linear combination of traits which maximizes the heritability and then tests the associations based on transformed traits. Canonical correlation analysis \citep{ferreira2009multivariate} finds the linear combination of traits maximizing the covariance between a SNP and all traits. It is equivalent to multivariate analysis of variance (MANOVA) when there is only one SNP \citep{van2013tates}. The third category relies on regression models. MultiPhen \citep{OReilly2012} regresses genotypes on phenotypes via proportional odds logistic model. As mentioned, GEE has been utilized to generate score test statistics in aSPU \citep{Zhang2014}. Besides linear models, kernel regression models (KMRs) also play a role in multiple-trait analysis, including multivariate kernel machine regression \citep{maity2012multivariate}, multi-trait sequence kernel association test (MSKAT) \citep{Wu2016} and Multi‐SKAT \citep{dutta2019multi}. \citet{davenport2018powerful} extended KMRs to multiple binary outcomes.

Currently, several limitations still exist in multiple-trait methods restricting their wide applications. First, many existing methods are unable to simultaneously analyze binary and continuous traits. For KMRs, multiple non-continuous traits can be both theoretically and computationally challenging and it is unclear how to integrate multiple different data types (e.g., multi-omics) \citep{larson2019review}. MANOVA does not apply to non-Normal traits. Numerous studies in GWASs are case-control studies and the incapability of processing binary traits greatly limits their application. Second, the methods may have inconsistent performance under various scenarios depending on the number of traits, the strength of correlation, and the number of true associations, which are largely unknown in practice. Hence, there is a demand for methods with robust performance regardless of scenarios. Third, although many methods claimed the capability of handling covariates and confounders, few of them conducted the relevant simulations or real-data applications to demonstrate the type I error control and performance. As a matter of fact, our simulation study indicates the claims of some methods are inaccurate (see section \ref{sec:results}).

There is another kind of methods focusing on GWASs summary statistics such as MTAG \citep{turley2018multi}, mashr \citep{urbut2019flexible}, and MTAR \citep{luo2020multi}. Despite their computational efficiency, working with the individual-level data might still be favored by researchers for several reasons \citep{dai2017igess}. The individual-level data have more information than summary statistics data and the summary statistics methods can be inaccurate when linkage disequilibrium is mishandled \citep{dai2019joint}. Moreover, sometimes it is necessary to use individual-level data rather than summary statistics, for example, image data. 

In this paper, We propose a multiple-trait adaptive Fisher's (MTAF) method for multiple traits based on adaptive Fisher's (AF) method \citep{Song2016}. It combines the single-trait p-values adaptively to aggregate the evidence and being adaptive means its estimation adapts to the specific condition in the data. A similar method named AFC was proposed by \citet{liang2016adaptive} for GWASs, but it did not allow for covariate adjustment. 

The remainder of this paper is organized as follows. In section \ref{sec:methods}, we elaborate the proposed method and introduce variations to tackle highly correlated traits. In section \ref{sec:results}, we evaluate the performance of MTAF using simulation and apply it to a real GWAS of substance addiction traits. In section \ref{sec:discussion}, we review the advantages and the limitations of the MTAF method and discuss our future work.

\section*{Materials and methods}
\label{sec:methods}

\begin{figure}
    \centering
    \begin{subfigure}[t]{0.45\textwidth}
        \centering
        \includegraphics[width=1.5\linewidth]{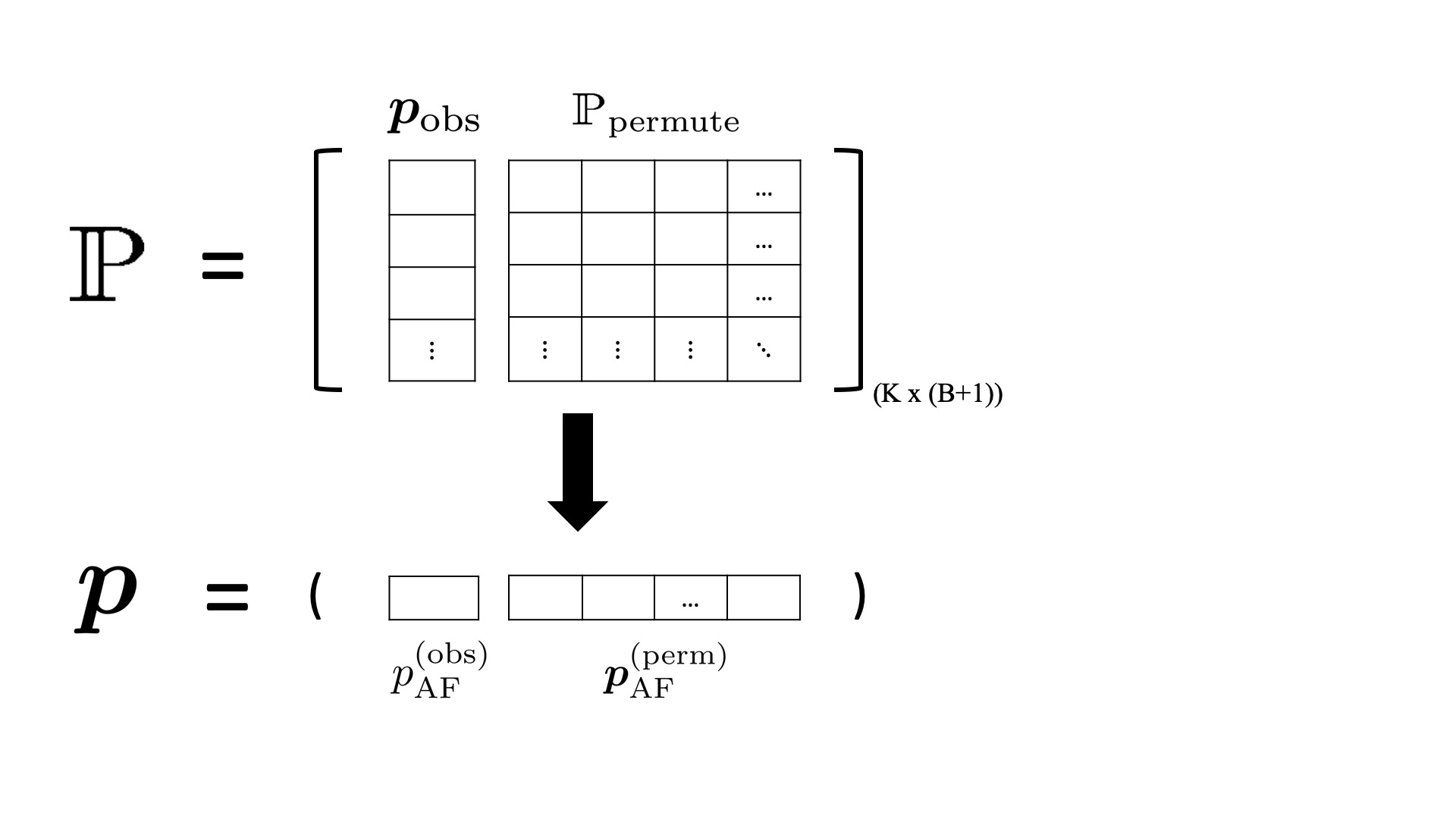} 
        \caption{AF combines p-values} 
        \label{fig:AF1}
    \end{subfigure}
    \hfill
    \begin{subfigure}[t]{0.45\textwidth}
        \centering
\begin{tikzpicture}[node distance=1.2cm]

\node (P_lower) [io] {$\mathbbm{P}_{lower}$};
\node (P_upper) [io, right of=P_lower, xshift=2cm] {$\mathbbm{P}_{upper}$};
\node (p_lower) [io,below of=P_lower] {$\bm{p}_{lower}$};
\node (p_upper) [io,below of=P_upper] {$\bm{p}_{upper}$};
\node (P_combo) [io,xshift=0.5cm,yshift=-0.5cm,below right of= p_lower] {$\mathbbm{P}_{combo}=[\bm{p}_{lower} \; \bm{p}_{upper}]$};
\node (p_combo) [io,below of=P_combo] {$\bm{p}_{combo}$};

\draw [arrow] (P_lower) -- node[anchor=west] {AF} (p_lower);
\draw [arrow] (P_upper) -- node[anchor=west] {AF} (p_upper);
\draw [arrow] (p_lower) -- node[anchor=west] {} (P_combo);
\draw [arrow] (p_upper) -- node[anchor=west] {} (P_combo);
\draw [arrow] (P_combo) -- node[anchor=west] {AF} (p_combo);

\end{tikzpicture}
\caption{$\text{AF}_{\text{1side}}$ combines one-sided p-values} \label{fig:AF2}
    \end{subfigure}

    \vspace{1cm}
    \begin{subfigure}[t]{\textwidth}
    \centering
\begin{tikzpicture}[node distance=1.2cm]

\node (data) [startstop] {data: $\bm{Y}$,$\bm{x}$};
\node (y_bin) [startstop, below of=data, xshift=-2.5cm] {$\bm{Y}_{bin}$,$\bm{x}$};
\node (y_cont) [startstop, below of=data,xshift=5cm] {$\bm{Y}_{cont}$,$\bm{x}$};
\node (score_bin) [process,below of=y_bin] {Score test};
\node (y_org) [startstop, below of=y_cont,xshift=-2.5cm] {$\bm{Y}_{original}$,$\bm{x}$};
\node (y_pca) [startstop, right of=y_org, xshift=2.5cm] {$\bm{Y}_{pca}$,$\bm{x}$};
\node (score_org) [process, below of=y_org] {Score test};
\node (score_pca) [process, below of=y_pca] {Score test};
\node (P_bin) [io, below of=score_bin] {$\mathbbm{P}_{lower}$, $\mathbbm{P}_{upper}$};
\node (P_org) [io, below of=score_org] {$\mathbbm{P}_{lower}$, $\mathbbm{P}_{upper}$};
\node (P_pca) [io, below of=score_pca] {$\mathbbm{P}_{pca}$};
\node (p_bin) [io, below of=P_bin,yshift=-3.6cm] {$\bm{p}_{bin}$};
\node (p_org) [io, below of=P_org] {$\bm{p}_{original}$};
\node (p_pca) [io, below of=P_pca] {$\bm{p}_{pca}$};
\node (P_cont) [io, below of=p_org, xshift=2cm] {$\mathbbm{P}_{cont}=[\bm{p}_{original}\:\: \bm{p}_{pca}]$};
\node (p_cont) [io, below of=P_cont] {$\bm{p}_{cont}$};
\node (P_mix) [io, below of=p_cont, xshift=-3cm] {$\mathbbm{P}_{mix}=[\bm{p}_{bin}\:\: \bm{p}_{cont}]$};
\node (p_mix) [io, below of=P_mix] {$\bm{p}_{mix}$};

\draw [arrow] (data) -- node[left=0.5em] {binary} (y_bin);
\draw [arrow] (data) -- node[anchor=east] {continuous} (y_cont);
\draw [arrow] (y_cont) -- node[anchor=east] {} (y_org);
\draw [arrow] (y_cont) -- node[anchor=east] {PCA} (y_pca);
\draw [arrow] (y_bin) -- node[anchor=east] {permute} (score_bin);
\draw [arrow] (y_org) -- node[anchor=east] {permute} (score_org);
\draw [arrow] (y_pca) -- node[anchor=east] {permute} (score_pca);
\draw [arrow] (score_bin) -- node[anchor=east] {} (P_bin);
\draw [arrow] (score_org) -- node[anchor=east] {} (P_org);
\draw [arrow] (score_pca) -- node[anchor=east] {} (P_pca);
\draw [arrow] (P_bin) -- node[anchor=east] {$\text{AF}_{\text{1side}}$} (p_bin);
\draw [arrow] (P_org) -- node[anchor=east] {$\text{AF}_{\text{1side}}$} (p_org);
\draw [arrow] (P_pca) -- node[anchor=east] {AF} (p_pca);
\draw [arrow] (p_org) -- node[anchor=east] {} (P_cont);
\draw [arrow] (p_pca) -- node[anchor=east] {} (P_cont);
\draw [arrow] (P_cont) -- node[anchor=east] {AF} (p_cont);
\draw [arrow] (p_bin) -- node[anchor=east] {} (P_mix);
\draw [arrow] (p_cont) -- node[anchor=east] {} (P_mix);
\draw [arrow] (P_mix) -- node[anchor=east] {AF} (p_mix);

\end{tikzpicture}
        \caption{Workflow of analyzing mixed traits} \label{fig:AF3}
    \end{subfigure}
    \caption{Summary of methods under MTAF framework}
\end{figure}
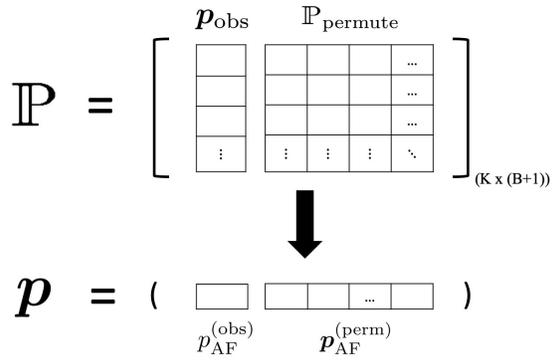
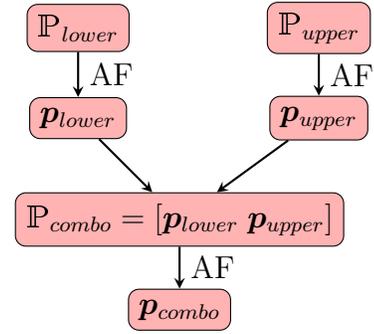
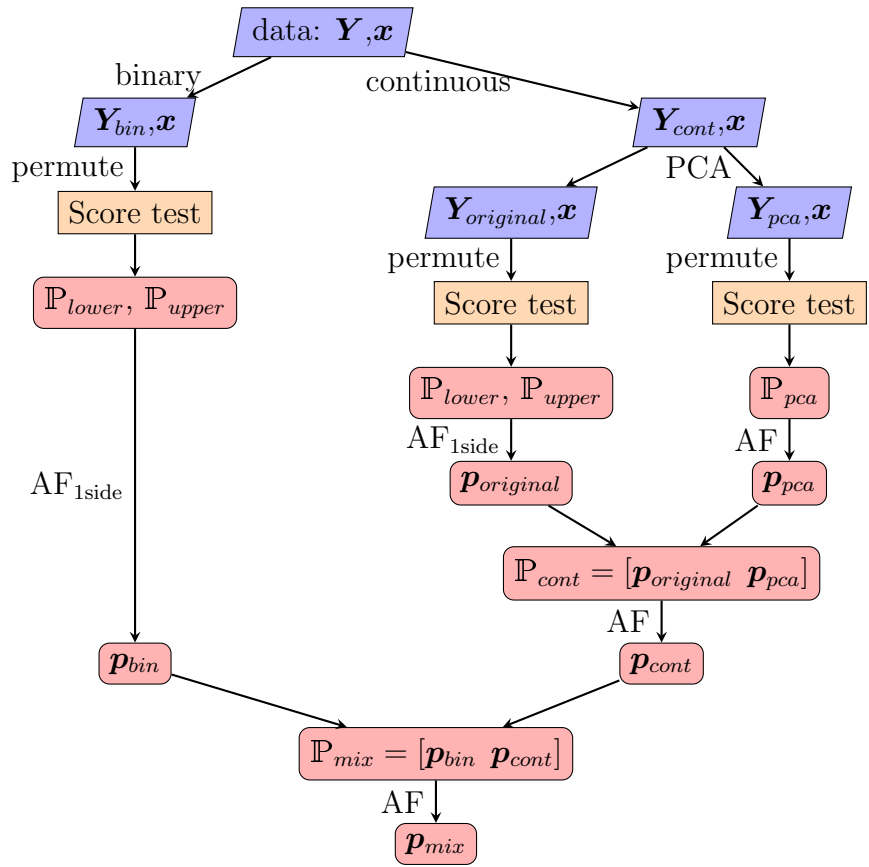

Suppose there are $n$ independent subjects. For each subject $i=1,\ldots,n$, there are $K$ traits $\bm{Y_i}=(y_{i1},\ldots,y_{iK})'$ and $\bm{Y_k}=(y_{1k},\ldots,y_{nk})'$, and $x_i \in \{0,1,2\}$ is the genotype of a SNP coded as the number of minor alleles in the subject $i$. $\bm{x} = (x_1,\ldots,x_n)$ is the vector of all subjects' genotypes for this SNP. In terms of real applications, we suppose each subject $i$ has $M$ covariates (e.g. sex and race) $z_{i1},\dots,z_{iM}$ and $\bm{Z}=\{z_{im}\}_{n \times M}$. After adjusting for covariates, We aim to test the associations between the SNP and $K$ traits under the null hypothesis $H_0$ that none of the $K$ traits associates with the SNP. To construct the test statistics, the MTAF method combines the marginal p-values from the single-trait score tests in an adaptive way. At last, because our test statistic has no closed form, we conduct permutations to get the empirical p-values.

\subsection*{Score Test}
Score test is one of the most popular test in GWASs of single traits, because it only requires fitting the null model thus is computationally inexpensive. Suppose for each SNP and subject $i$, a generalized linear model of the following form is assumed for the $k^{th}$ trait and the SNP with $M$ covariates:

\begin{equation*}
g_{k}(E(Y_{ik})) = x_{i} \cdot \beta_{k} + \sum_{m=1}^{M} z_{im} \cdot \alpha_{mk},
\label{eq:glm}
\end{equation*}
where $\beta_k$ is the effect of the SNP and $\alpha$'s are the effects of the M covariates on the $k^{th}$ trait. $g_k(\cdot)$ is the link function, which is identity function for continuous traits and logit function for binary traits. Different link functions make it possible to allow for both continuous traits and binary traits. Under $H_0: \beta_{k} = 0$, the test statistic of score test for the SNP is:

\begin{equation*}
U_k = \sum_{i=1}^n (x_i-\hat{x}_i)(Y_{ik}-\hat{Y}_{ik}),
\label{eq:score}
\end{equation*}
where $\hat{x}_i$ is an estimate of $x_i$ and $\hat{Y}_{ik}$ is an estimate of $Y_{ik}$. Denote $\bm{V}$ as Fisher information matrix which is the covariance matrix of $\bm{U}=(U_1,\ldots,U_k)$ and $\Var(U_k|H_0)=V_{kk}$ where $V_{kk}$ is the $k^{th}$ diagonal element of $\bm{V}$. Asymptotically, we have $U_k/\sqrt{V_{kk}} \sim  \mathcal{N}(0,1)$ and then the p-values (either one-sided or two-sided) can be generated. We get the p-values of the single-trait score tests computationally by R package \textbf{statmod} \citep{RJ-2016-024}.

\subsection*{MTAF Method}
Denote $p_1,...,p_K$ as the p-values of score tests between the K traits and the SNP. Let $S_k = - \sum_{i=1}^k{\log{p_{(i)}}}$ where $p_{(i)}$ is the $i^{th}$ smallest p-value and it is the sum of first $k$ smallest p-values. Then the p-value of $s_k$ is $p_{s_k} = P( S_k \ge s_k )$ where $s_k$ is the observed value of $S_k$. In practice, this p-value can be obtained by permutation.

The proposed test statistic of the MTAF method  is

\begin{equation*}
T_{MTAF} = \min_{1 \le k \le K} p_{s_k}.
\end{equation*}

\subsection*{Permutation test}



Since it is intractable to get the p-value of the test statistic analytically, we turn to apply the permutation procedure to get the empirical p-values. We intend to test the conditional independence between the SNP and the traits given the covariates, i.e.,$\bm{Y} \perp \bm{x} \mid \bm{Z}$. The permutation procedure should break the associations between $\bm{Y}$ and $\bm{x}$ while preserving the associations between $\bm{Y}$ and $\bm{Z}$ and between $\bm{x}$ and $\bm{Z}$. Simply permuting the genotype $\bm{x}$ leads to inflated type I error rate, because the correlation between the genotype and covariates are destroyed. Following \citet{potter2005permutation} and \citet{werft2010glmperm}, we permute residuals of regressions of $\bm{x}$ on $\bm{Z}$ for generalized regression models. In our method, we first regress the genotype on the covariates, then we permute the residuals derived from the regression. We replace the original genotype with the permuted residuals to perform score tests for the permuted data. It should be noted that even when no covariate is explicitly included in the mode, we still have a constant as our covariate.

Specifically, we denote the vector of residuals of regressing $\bm{x}$ on $\bm{Z}$ as $\bm{e_{x}}$ and permute it for B times. In the $b^{th}$ permutation, we regress $\bm{Y_{k}}$ on $\bm{e_{x}^{(b)}}$ and get the score tests p-values $p_{k}^{(b)}$ for the coefficient of $\bm{e_{x}^{(b)}}$. After B permutations, we get a $(B+1) \times K$ matrix $\mathbbm{P}=\{p_{k}^{(b)} \}$. Each element $p_{k}^{(b)}$ is the p-value measuring the $k^{th}$ trait in the $b^{th}$ permutation for $1 \leq b \leq B$ and $p_{k}^{(0)}$ is the observed p-value. Based on $\mathbbm{P}$, we can construct the MTAF method's test statistics for both the observed data and permuted data.

For the matrix $\mathbbm{P}$, we can calculate the empirical p-values of the MTAF method for the observed data and permuted data with the following steps: 

Suppose we have a $(B+1) \times K$ matrix of p-values $\mathbbm{P}$.

\begin{enumerate}

\item For each row $b \in \{ 0,1,...,B \}$, we calculate $s_k^{(b)}$ and 

\begin{equation*}
p_{s_k}^{(b)} = \frac{1}{B+1} \sum_{j=0}^B \mathbbm{1}{\{s_k^{(j)} \geq s_k^{(b)}\}},
\end{equation*}

where $\mathbbm{1}$ is the indicator function.

\item Then we can get a vector $\bm{t}=(t_{MTAF}^{(0)},t_{MTAF}^{(1)},\dots,t_{MTAF}^{(B)})$ where $t_{MTAF}^{(b)} = \min_{1 \le k \le K} p_{s_k}^{(b)}$.

\item The p-values of MTAF test statistics are approximated by

\begin{equation*}
    p_{MTAF}^{(b)} = \frac{1}{B+1} \sum_{j=0}^B \mathbbm{1}{\{ t_{MTAF}^{(j)} \leq t_{MTAF}^{(b)} \}},
\label{eq:approx_pval}
\end{equation*}
where $p_{MTAF}^{(b)}$ is the empirical p-value of the MTAF method for the permuted data for $1 \leq b \leq B$ and $p_{MTAF}^{(0)}$ is the empirical p-value for the observed data.
\end{enumerate}

To simplify the following discussion of the variation of MTAF method, we define the steps above as an AF operator $AF\{\cdot\}$ mapping $\mathbbm{P}$ to $\bm{p} = (p_{MTAF}^{(0)},p_{MTAF}^{(1)},\dots,p_{MTAF}^{(B)})$ (Figure \ref{fig:AF1}).

\subsection*{Combination of One Sided P-values}

In practice, traits of a complex disease tend to be positively correlated intrinsically. Or, according to the prior knowledge, we can manually change the direction of effects to make them in the same direction. In the situation that effects are in the same direction, combining one-sided p-values aggregates evidence for the effects which tend to have the same signal and enjoys higher statistical power than combining two-sided p-values. Therefore, we recommend always combine one-sided p-values when it is appropriate. We separately combine the lower-tail p-values and the upper-tail p-values, and then unify these two results using another round of MTAF permutation. Specifically, we get $\bm{p}_{lower} = AF\{\mathbbm{P}_{lower}\}$ and $\bm{p}_{upper} = AF\{\mathbbm{P}_{upper}\}$ and then the empirical p-value of the observed data is the first element of $\bm{p}_{combo} = AF\{[\bm{p}_{lower} \ \bm{p}_{upper}] \}$ (Figure \ref{fig:AF2}).

\subsection*{PCA of continuous traits}
  
Alternatively, a SNP may not have strong associations with observed traits but hidden components, which can be difficult to detect those associations. PCA is widely used to reduce dimensions and it generates orthogonal linear combinations of variables maximizing the variability. We introduce PCA into the MTAF method with the purpose of uncovering the hidden components. In the MTAF method, PCA generates $K$ independent principal components and we detect the associations between the SNP and  the principal components. Specifically, for continuous traits, we first regress $\bm{Y_k}$ on the covariates $\bm{Z}$ and denote the residuals $\bm{e_{k}}$ for $1 \leq k \leq K$. PCA conducted on $\bm{e_{1}},\ldots,\bm{e_{K}}$ leads to $K$ principal components which substitute for $\bm{y_{1}},\ldots,\bm{y_{K}}$. In the simulation study, the power of the MTAF method increases dramatically given correlated traits after applying PCA. Unlike its common use in practice that selects only several top principal components, it claims that using all principal components can have greater power \citep{aschard2014maximizing}. Therefore, the MTAF method keeps all principal components and it proves to be powerful. The MTAF method itself is powerful when signals are sparse, but usually the number of traits truly associated with the SNP and underlying correlation structure are unknown. Hence, when analyze continuous traits, initially we apply the original MTAF method and the MTAF method with PCA respectively, then combine the results from the two to get the final p-value. Specifically, we have $\bm{p}_{original}=AF(\mathbbm{P}_{original})$ and $\bm{p}_{pca}=AF(\mathbbm{P}_{pca})$. Then combine two vectors $\bm{p}_{original}$ and $\bm{p}_{pca}$ into a matrix $\mathbbm{P}_{continuous}$. At last, we have $\bm{p}_{continuous}=AF(\mathbbm{P}_{continuous})$ and the p-value is the first element of $\bm{p}_{continuous}$. To process a mixture of binary traits and continuous traits, we first apply the MTAF method to binary traits to get $\bm{p}_{binary}$ and then get $\bm{p}_{continuous}$ by the procedure above. Then we combine two p-value vectors to get $\mathbbm{P}_{mix} = [\bm{p}_{binary} \ \bm{p}_{continuous}]$ and the empirical p-value is the first element of $\bm{p}_{mix}=AF(\mathbbm{P}_{mix})$. The workflow of analyzing mixed traits is illustrated in Figure \ref{fig:AF3}.

In the MTAF method, PCA is used to handle continuous traits rather than binary traits. Although some literature refers to generalized PCA \citep{landgraf2019generalized}, our method only applies PCA to continuous traits at this moment.

\subsection*{Simulation Setup}

To evaluate the performance of the MTAF method, we conduct a simulation study. In each dataset, we simulate $1000$ subjects based on various parameters such as the number and types of traits, as well as the proportion of traits associated with the genotype and the strength of the association. We consider $10$, $50$, and $100$ traits and we simulate three scenarios: continuous traits only, binary traits only, and a mixture of the two. We assume a compound symmetry (CS) structure underlying traits with either weak correlation ($\rho=0.3$) or strong correlation ($\rho=0.6$). For the proportion of associated traits, we define the sparse scenarios as when $2\%$ of the traits are truly associated with the SNP, and the dense scenarios as when $20\%$ are associated. However, when there are only $10$ traits, we set the number of associated traits being 1 and 4 for the sparse and dense scenarios respectively. The detailed simulation steps are listed below.

First, we simulate the genotypes of the SNP. Since we focus on the association between single common SNPs and multiple traits, we only simulate one SNP genotype $x_i \in \{0,1,2\}$ for each subject $i$, such that $x_i \sim \textrm{Bin} (2, 0.3)$, where $0.3$ is the minor allele frequency (MAF) of the simulated SNP.

Next, the traits for each subject $\bm{Y_i}=(Y_{i1},\ldots,Y_{iK})'$ are simulated via a linear model:

\begin{equation}
Y_i = x_i\bm{\beta} + \bm{\epsilon}_i,
\label{eq:sim_1}
\end{equation}
where $\bm{\beta} = (\beta_1,\ldots,\beta_K  )$ are the coefficients. The non-zero $\beta_k$'s are drawn from independent uniform distributions and we select the parameters of the uniform distributions to make the differences among methods obvious. $\bm{\epsilon}_i$'s are independently drawn from a multivariate Gaussian distribution $N(\mathbf{0},\bm{\Sigma})$. To simulate correlated traits, $\bm{\Sigma}$ is simulated such that the variances are sampled independently from an inverse gamma distributions $\textrm{Inv-Gamma}(4, 4)$ and the corresponding correlation matrix is CS with correlation $\rho$.

In addition, we consider simulation scenarios with two binary covariates $Z_{i1}$ and $Z_{i2}$ to investigate the performance of the MTAF method in the presence of confounders, such as gender and race in the real datasets. These covariates are simulated by dichotomizing underlying covariates that are linearly associated with the genotype.  Specifically, $Z_{i1}$ is simulated by dichotomizing $x_i \eta_1 + \omega_{i1}$, and $Z_{i2}$ are simulated by dichotomizing $x_i \eta_2 + \omega_{i2}$, where $\eta_1$ and $\eta_2$ are randomly drawn from uniform distributions $U(0.5,1)$, and $\omega_{i1}$ and $\omega_{i2}$ follow $\mathcal{N}(0,1)$. Then, we label the values greater than medians ``1", otherwise ``0". $\bm{Y_i}$ is simulated based on a linear model conditional on both the genotype and the covariates:

\begin{equation}
\bm{Y_i} = x_i \bm{\beta} + \bm{Z_i} \bm{\Gamma} + \bm{\epsilon}_i,
\label{eq:sim_2}
\end{equation}
where $\bm{Z_i} = (Z_{i1},Z_{i2})$ and $\bm{\Gamma}_{K \times 2}$ has coefficients drawn from iid uniform distribution $U(0.5,1)$.

To simulate binary traits, we first simulate the log-odds of $Y_{ik}=1$ by replacing the corresponding $Y_{ik}$ with $\textrm{logit}(E(Y_{ik}))$ in \ref{eq:sim_1} and  \ref{eq:sim_2} and then we draw the binary traits based on the simulated odds. 

To evaluate the performance of the MTAF method, competitor methods including MSKAT, aSPU, TATES, MANOVA, MultiPhen, and minP are also applied on the simulated datasets for comparison.

\subsection*{Data availability}

The authors state that all data necessary for confirming the conclusions are represented fully within the article. The SAGE data was downloaded from the dbGAP using accession number phs000092.v1.p1. The R software package for the MTAF method and our simulation code are available at \url{https://github.com/songbiostat/MTAF}.

\section*{Results}
\label{sec:results}

\subsection*{Simulation Results}

\subsubsection*{Type I Error Rate}

First, to assess whether the MTAF method and other methods can appropriately control type I error at the nominal level, we perform simulations under the null hypothesis where $\beta_1=\dots=\beta_K=0$. The empirical p-values were calculated based on $1,000$ permutations and the type I error rate is evaluated at the $0.05$ nominal level. In addition to aSPU with default independent structure, we evaluated aSPU equipped with exchangeable correlation structure (aSPU-ex).

Table \ref{tab:typeI_cont} shows that, when all traits were continuous, the empirical Type I error of most methods were well controlled allowing for different number of traits and strength of correlation. We found that MultiPhen had inflated Type I error, especially after adding covariates into the models. Thus, we decided not to include MultiPhen in the corresponding simulation studies. The similar phenomenon was reported by Konigorski et al.\citep{konigorski2020powerful} that MultiPhen led to inflated or highly inflated type I errors and they did not include the method in the power study. Table \ref{tab:typeI_bin} and \ref{tab:typeI_mix} show that type I error were well controlled for the compared methods when all traits are binary, or when the traits are half binary and half continuous. Please be noted that only methods that can be applied on binary or mixed trait scenarios are included in tables \ref{tab:typeI_bin} and \ref{tab:typeI_mix}.


\begin{table}[htbp]
\caption{\bf Type I error: continuous traits} 
\centering 
\scalebox{0.8}{
\begin{tabular}{c c c c cccccccc} 
\hline 
\# Covariates & Correlation & \# Traits & MTAF & MSKAT & aSPU  &aSPU-ex &MultiPhen & TATES & MANOVA  & minP
\\ [0.5ex]
\hline 
 0 & 0.3 & 10 &0.042	 &0.048	&0.041	&0.049	&0.049	&0.045	&0.050	&0.049 \\[-0.5ex] 
 & & 50 &0.044	&0.038	&0.047	&0.039	&0.038	&0.042	&0.045	&0.049 \\[-0.5ex]   
 & & 100 &0.049	&0.040	&0.054	&0.051	&0.041	&0.050	&0.049	&0.052 \\[0.5ex]   

 &  0.6 & 10 &0.045	&0.048	&0.041	&0.049	&0.049	&0.035	&0.050	&0.039 \\[-0.5ex] 
 & & 50 &0.041	&0.038	&0.050	&0.039	&0.038	&0.026	&0.045	&0.043 \\[-0.5ex]   
 & & 100 &0.051	&0.040	&0.061	&0.047	&0.041	&0.044	&0.049	&0.054 \\[1ex]   

2 & 0.3 & 10 &0.051	&0.049	&0.039	&0.042	&0.066	&0.046	&-	&0.046 \\[-0.5ex] 
& & 50 &0.049	&0.064	&0.040	&0.041	&0.097	&0.045	&-	&0.046 \\[-0.5ex]   
& & 100 &0.047 	&0.042	&0.056	&0.058	&0.146	&0.043	&-	&0.046 \\[0.5ex]   

& 0.6 & 10 &0.048	&0.048	&0.043	&0.050	&0.059	&0.044	&-	&0.045 \\[-0.5ex] 
& & 50 &0.039	&0.064	&0.034	&0.038	&0.097	&0.027	&-	&0.038 \\[-0.5ex]   
& & 100 &0.043 	&0.040	&0.033	&0.042	&0.138	&0.029	&-	&0.049 \\[0ex]   

\hline

\hline 
\end{tabular}
}
\label{tab:typeI_cont}
\end{table}


\begin{table}[htbp]
\caption{\bf Type I error: binary traits} 
\centering 
\scalebox{1}{
\begin{tabular}{c c c c cccccc} 
\hline 
\# Covariates & Correlation & \# Traits & MTAF &aSPU &aSPU-ex & MultiPhen &TATES  & minP
\\ [0.5ex]
\hline 
 0 & 0.3 & 10 &0.050	 &0.055	&0.056	&0.066	&0.054	&0.055 \\[-0.5ex] 
 & & 50 &0.050 &0.051	&0.040	&0.072	&0.055	&0.052 \\[0.5ex]   

 &  0.6 & 10 &0.044	&0.048	&0.055	&0.059	&0.048	&0.046 \\[-0.5ex] 
 & & 50 &0.051	&0.046	&0.049	&0.077	&0.055	&0.054 \\[1ex]   

 2 & 0.3 & 10 &0.039 &0.042	&0.043	&0.526	&0.047	&0.049 \\[-0.5ex] 
 & & 50 &0.035	&0.034	&0.037	&0.710	&0.044	&0.043 \\[0.5ex]   

 &  0.6 & 10 &0.043	&0.042	&0.039	&0.429	&0.044	&0.040 \\[-0.5ex] 
 & & 50 &0.038	&0.038	&0.044	&0.606	&0.049	&0.052 \\[0ex]   
 
\hline 
\end{tabular}
}
\label{tab:typeI_bin}
\end{table}


\begin{table}[htbp]
\caption{\bf Type I error: mixed traits with covariates} 
\centering 
\scalebox{1}{
\begin{tabular}{c c c ccc} 
\hline 
Correlation & \# Traits & MTAF &MultiPhen & TATES  & minP
\\ [0.5ex]
\hline 
0.3 & 10 &0.054	&0.886	&0.059	&0.058	 \\[-0.5ex] 
& 50 &0.043	&0.982	&0.056	&0.055 \\[0.5ex]   

0.6 & 10 &0.049	&0.878	&0.053	&0.051 \\[-0.5ex] 
& 50 &0.040	&0.996	&0.041	&0.045\\[0ex]   

\hline 
\end{tabular}
}
\label{tab:typeI_mix}
\end{table}

\subsubsection*{Statistical Power}

The power of these compared methods were evaluated under different scenarios at significance level of $0.05$. The effect sizes for the associated traits were randomly drawn from uniform distributions. We include the original MTAF method ($\textrm{MTAF}_{\textrm{original}}$) and its PCA expansion ($\textrm{MTAF}_{\textrm{PCA}}$). The results show that the statistical power under dense scenarios was greatly improved by introducing PCA. Table \ref{tab:power_cont_noz} summarizes the power of the compared methods for continuous traits without covariates. We observe that when signals were sparse, the MTAF method was the most powerful method or performed similar to the most powerful method. On the other hand, when the signals were dense, MSKAT, MANOVA, and MultiPhen were the powerful methods.  Although the MTAF method had a slightly lower power, the performance of the MTAF method was close to these methods. Table \ref{tab:power_cont_z} shows the results for continuous traits with two covariates. It shows that when confounders were included, only the MTAF method and MSKAT managed to preserve their power in both sparse and dense scenarios, while the performance of other methods deteriorated for the dense signal scenarios, especially when the number of traits got large. Table \ref{tab:power_bin_noz} shows the results for binary traits without covariates. Under this scenario, aSPU and aSPU-ex outperformed other methods. The MTAF method was slightly less powerful than aSPU methods, while TATES and minP performed well with sparse signal, but significantly underperforms with dense signals.  In Table \ref{tab:power_bin_z}, we find that with two covariates,  performance difference between the MTAF method and the aSPU methods diminished, and their powers were close in most simulations settings. Table \ref{tab:power_mix} shows the results for mixed traits with two covariates.  It should be noted that only four methods allow for mixture of binary and continuous traits, including MultiPhen, the MTAF method, TATES, and minP. Whereas we did not include MultiPhen in our comparison because it fails to control type I error as shown previously. According to the results, the MTAF method outperformed TATES and minP regardless of the number of traits, the strength of correlation, or the proportional of signals. In summary, MTAF was robustly one of the most powerful methods in all the simulation settings with various number of traits, strength of correlation, and proportion of signals, for both continuous traits and binary traits (or their mixture), with or without confounding covariates.


\begin{table*}[!htbp]
\caption{\bf Power: continuous traits without covariates} 
\centering 
\scalebox{0.6}{
\begin{tabular}{c c c c cccccccccc} 
\hline 
Sparsity & Correlation & \# Traits  & Effect Size &$\textrm{MTAF}_{\textrm{PCA}}$ &$\textrm{MTAF}_{\textrm{original}}$ &MTAF	&MSKAT	&aSPU	&aSPU-ex 	&MultiPhen	&TATES	&MANOVA 	&minP
\\ [0.5ex]

\hline 
sparse & 0.3 & 10 &U(0.15,0.25) &0.762 &0.791 &0.793	&0.796	&0.563	&0.702	&0.796	&0.785	&0.798	&0.788 \\[-0.5ex] 
& & 50 & U(0.2,0.4) &0.797	&0.922 &0.916	&0.825	&0.688	&0.807	&0.827	&0.904	&0.836	&0.906 \\[-0.5ex]   
& & 100 & U(0.15,0.3) &0.770	&0.924 &0.919	&0.843	&0.400	&0.665	&0.845	&0.895	&0.852	&0.894 \\[0.5ex]
& 0.6 & 10 &U(0.15,0.25) &0.907	&0.858 &0.905	&0.918	&0.590	&0.859	&0.918	&0.799	&0.918	&0.907 \\[-0.5ex] 
& & 50 & U(0.2,0.4) &0.916	&0.957 &0.949	&0.944	&0.730	&0.926	&0.944	&0.908	&0.950	&0.924 \\[-0.5ex]   
& & 100 & U(0.15,0.3) &0.935	&0.948 &0.968	&0.960	&0.413	&0.833	&0.960	&0.886	&0.966	&0.914 \\[1ex]

dense & 0.3 & 10 &U(0.05,0.15) &0.769	&0.690 &0.765	&0.784	&0.390	&0.507	&0.784	&0.651	&0.786	&0.639 \\[-0.5ex] 
& & 50 & U(0.05,0.12) &0.808	&0.629 &0.812	&0.865	&0.134	&0.422	&0.866	&0.534	&0.873	&0.551\\[-0.5ex]   
& & 100 & U(0.02,0.1) &0.615	&0.418 &0.637	&0.716	&0.082	&0.266	&0.716	&0.334	&0.742	&0.343 \\[0.5ex]
& 0.6 & 10 &U(0.05,0.15) &0.92	&0.729 &0.907	&0.933	&0.301	&0.701	&0.933	&0.627	&0.933	&0.641 \\[-0.5ex] 
& & 50 & U(0.05,0.12) &0.969	&0.654 &0.964	&0.987	&0.119	&0.640	&0.987	&0.457	&0.988	&0.526 \\[-0.5ex]   
& & 100 & U(0.02,0.1) &0.929	&0.437 &0.916	&0.970	&0.074	&0.338	&0.970	&0.273	&0.974	&0.346 \\[0ex]
 
\hline 
\end{tabular}
}
\label{tab:power_cont_noz}
\end{table*}


\begin{table}[htbp]
\caption{\bf Power: continuous traits with covariates} 
\centering 
\scalebox{0.7}{
\begin{tabular}{c c c c cccccccc} 
\hline 
Sparsity & Correlation & \# Traits  & Effect Size &$\textrm{MTAF}_{\textrm{PCA}}$ &$\textrm{MTAF}_{\textrm{original}}$ &MTAF	&MSKAT	&aSPU &aSPU-ex	&TATES &minP
\\ [0.5ex]

\hline 

sparse & 0.3 & 10 &U(0.15,0.3) &0.763	&0.787 &0.803	&0.795	&0.602	&0.715	&0.773	&0.779 \\[-0.5ex] 
& & 50 & U(0.2,0.4) &0.684	&0.889 &0.871	&0.757	&0.578	&0.722	&0.871	&0.876 \\[-0.5ex]   
& & 100 & U(0.15,0.3) &0.646	&0.872 &0.862	&0.754	&0.273	&0.543	&0.835	&0.840 \\[0.5ex]

& 0.6 & 10 &U(0.15,0.3) &0.908	&0.862 &0.900	&0.920	&0.620	&0.865	&0.778	&0.794 \\[-0.5ex] 
& & 50 & U(0.2,0.4) &0.877	&0.935 &0.939	&0.917	&0.619	&0.878	&0.880	&0.891 \\[-0.5ex]   
& & 100 & U(0.15,0.3) &0.889	&0.922 &0.948	&0.934	&0.307	&0.739	&0.828	&0.859 \\[1ex]

dense & 0.3 & 10 &U(0.05,0.2) &0.833	&0.792 &0.842	&0.875	&0.506	&0.649	&0.759	&0.748 \\[-0.5ex] 
& & 50 &U(0.05,0.13) &0.685	&0.517 &0.695	&0.775	&0.123	&0.387	&0.468	&0.470\\[-0.5ex]   
& & 100 &U(0.03,0.12) &0.460	&0.301 &0.478	&0.560	&0.064	&0.211	&0.258	&0.268 \\[0.5ex]

& 0.6 & 10 &U(0.05,0.2) &0.957	&0.833 &0.947	&0.963	&0.422	&0.821	&0.732	&0.737 \\[-0.5ex] 
& & 50 &U(0.05,0.13) &0.934	&0.544 &0.928	&0.963	&0.105	&0.567	&0.400	&0.457\\[-0.5ex]   
& & 100 &U(0.03,0.12) &0.779	&0.292 &0.756	&0.862	&0.057	&0.247	&0.192	&0.269 \\[0ex]

\hline 
\end{tabular}
}
\label{tab:power_cont_z}
\end{table}


\begin{table}[htbp]
\caption{\bf Power: binary traits without covariates} 
\centering 
\scalebox{0.7}{
\begin{tabular}{c c c c ccccc} 
\hline 
Sparsity & Correlation & \# Traits  & Effect Size &MTAF	&aSPU &aSPU-ex		&TATES &minP
\\ [0.5ex]

\hline 
sparse & 0.3 & 10 &U(0.4,0.6) &0.747	&0.837	&0.839	&0.805	&0.805 \\[-0.5ex] 
& & 50 &U(0.6,0.8) &0.891	&0.957	&0.959	&0.930	&0.934 \\[0.5ex]   

& 0.6 & 10 &U(0.4,0.6) &0.773	&0.84	&0.859	&0.804	&0.801\\[-0.5ex] 
& & 50 &U(0.6,0.8) &0.912	&0.960	&0.974	&0.934	&0.931 \\[1ex]   

dense & 0.3 & 10 &U(0.2,0.3) &0.712	&0.738	&0.723	&0.547	&0.554 \\[-0.5ex] 
& & 50 &U(0.15,0.3) &0.667	&0.734	&0.749	&0.473	&0.46 \\[0.5ex]   

& 0.6 & 10 &U(0.2,0.3) &0.684	&0.701	&0.691	&0.569	&0.573 \\[-0.5ex] 
& & 50 &U(0.15,0.3) &0.580	&0.619	&0.754	&0.459	&0.454 \\[0ex]   

\hline 
\end{tabular}
}
\label{tab:power_bin_noz}
\end{table}


\begin{table}[htbp]
\caption{\bf Power: binary traits with covariates} 
\centering 
\scalebox{0.7}{
\begin{tabular}{c c c c ccccc} 
\hline 
Sparsity & Correlation & \# Traits  & Effect Size &MTAF	&aSPU	&aSPU-ex 	&TATES &minP
\\ [0.5ex]

\hline 
sparse & 0.3 & 10 &U(0.4,0.6) &0.702	&0.704	&0.714	&0.745	&0.739 \\[-0.5ex] 
& & 50 &U(0.5,0.7) &0.745	&0.725	&0.753	&0.800	&0.805 \\[0.5ex]   

& 0.6 & 10 &U(0.4,0.6) &0.718	&0.712	&0.747	&0.737	&0.740 \\[-0.5ex] 
& & 50 &U(0.5,0.7) &0.766	&0.738	&0.792	&0.791	&0.787 \\[1ex]   

dense & 0.3 & 10 &U(0.2,0.3) &0.610	&0.589	&0.570	&0.509	&0.491 \\[-0.5ex] 
& & 50 &U(0.2,0.35) &0.782	&0.783	&0.797	&0.637	&0.608 \\[0.5ex]   

& 0.6 & 10 &U(0.2,0.3) &0.548	&0.534	&0.525	&0.490	&0.467 \\[-0.5ex] 
& & 50 &U(0.2,0.35) &0.725	&0.683	&0.806	&0.608	&0.589 \\[0ex]   

\hline 
\end{tabular}
}
\label{tab:power_bin_z}
\end{table}


\begin{table}[htbp]
\caption{\bf Power: mixed traits with covariates and dense signals} 
\centering 
\scalebox{0.8}{
\begin{tabular}{c c c c ccc} 
\hline 
Correlation & \# Traits  & Effect Size &MTAF &TATES &minP
\\ [0.5ex]

\hline 

0.3 & 10 &U(0.05,0.3) &0.844 	&0.805	&0.804	\\[-0.5ex] 
& 50 &U(0.05,0.25) &0.929	&0.846	&0.852	 \\[0.5ex]   

0.6 & 10 &U(0.05,0.3) &0.897	&0.798	&0.795	\\[-0.5ex] 
& 50 &U(0.05,0.25) &0.986	&0.827	&0.835	 \\[0ex]   

\hline 
\end{tabular}
}
\label{tab:power_mix}
\end{table}

\subsection*{The Study of Addiction: Genetics and Environment (SAGE)}

To further demonstrate the usage of the proposed method in real studies, we applied MTAF to The Study of Addiction: Genetics and Environment (SAGE) \citep{bierut2010genome} data from the database of Genotypes and Phenotypes (dbGaP) \citep{mailman2007ncbi}, \url{http://www.ncbi.nlm.nih.gov/projects/gap/cgi-bin/study.cgi?study_id=phs000092.v1.p1}. SAGE is a case-control GWAS of addiction with unrelated individuals where cases are defined as individuals with Diagnostic and Statistical Manual of Mental Disorders 4th Edition (DSM-IV) alcohol dependence (lifetime) and potentially other illicit drug dependence. The individuals were selected from three large studies including the Collaborative Study on the Genetics of Alcoholism (COGA), the Family Study of Cocaine Dependence (FSCD), and the Collaborative Genetic Study of Nicotine Dependence (COGEND).

From dbGaP, We downloaded the data of $3,847$ individuals who consented to provide their data for health research. Quality control was performed using PLINK 1.9 \citep{purcell2007plink}. We filtered data based on the genotyping rate (0.01), missingness (0.01), minor allele frequency (0.01), and Hardy-Weinberg equilibrium (p-value $\le$ 0.01). After the filtering, we ended up with $3,557$ individuals and $560,218$ SNPs. In order to detect SNPs associated with addiction, we selected 18 traits (summarized in Table \ref{tab:sage_variables}) that account for the addiction to alcohol, nicotine, marijuana, cocaine, opiates, and any other drug. Among these traits, 12 are binary and 6 are continuous. In addition, gender and race (black or white) are included in our analysis as confounding covariates. 


\begin{table}[htbp]
\caption{\bf SAGE data variables summary} 
\centering 
\scalebox{0.75}{
\begin{tabular}{c c c} 
\hline 
& variable name  & description 

\\ [0.5ex]

\hline 

& nic\_sx\_tot & number of nicotine symptoms endorsed  \\ [0.5ex]

& nic\_sx1	& tolerance to nicotine	\\ [0.5ex]

& nic\_sx2	& withdrawal from nicotine	\\ [0.5ex]
	
& cig\_daily & Has participant ever smoked cigarettes daily for a month or more?	\\ [0.5ex]

& mj\_sx\_tot & number of marijuana dependence symptoms endorsed  \\ [0.5ex]

& mj\_sx1 & tolerance to marijuana  \\ [0.5ex]

& mj\_sx2 & withdrawal to marijuana  \\ [0.5ex]

& coc\_sx\_tot & number of cocaine dependence symptoms endorsed  \\ [0.5ex]

& coc\_sx1 & tolerance to cocaine  \\ [0.5ex]

& coc\_sx2 & withdrawal to cocaine  \\ [0.5ex]

& op\_sx\_tot & number of opiates dependence symptoms endorsed  \\ [0.5ex]

& op\_sx1 & tolerance to opiates  \\ [0.5ex]

& op\_sx2 & withdrawal to opiates  \\ [0.5ex]

& alc\_sx\_tot & number of alcohol dependence symptoms endorsed   \\ [0.5ex]

& alc\_sx1 & tolerance to opiates  \\ [0.5ex]

& alc\_sx2 & withdrawal to opiates  \\ [0.5ex]

& max\_drinks & largest number of alcoholic drinks consumed in 24 hours   \\ [0.5ex]

& ever\_oth	& Has participant ever used drugs other than marijuana, cocaine or opiates?	\\ [0.5ex]
 
\hline 
\end{tabular}
}
\label{tab:sage_variables}
\end{table}


\begin{table}[htbp]
\caption{\bf Correlation of continuous traits in SAGE data} 
\centering 
\scalebox{0.75}{
\begin{tabular}{c c c c c c c} 
\hline 
& nicotine  & marijuana & cocaine & alcohol & opiate & max drinks

\\ [0.5ex]

\hline 

nicotine &1 &0.390  &0.392  &0.534 &0.218 &0.345 \\ [0.5ex]

marijuana &0.390 &1  &0.534    &0.472 &0.331 &0.287 \\ [0.5ex]

cocaine &0.392 &0.534  &1 &0.519 &0.377 &0.346 \\ [0.5ex]

alcohol &0.537 &0.472  &0.519 &1 &0.298 &0.574  \\ [0.5ex]

opiate &0.218 &0.331  &0.377 &0.298 &1 &0.175 \\ [0.5ex]

max drinks &0.350 &0.287  &0.346  &0.574 &0.175 &1  \\ [0.5ex]

\hline 
\end{tabular}
}
\label{tab:sage_corr}
\end{table}

\begin{figure}[htbp!]
\centering
\includegraphics[width=\linewidth]{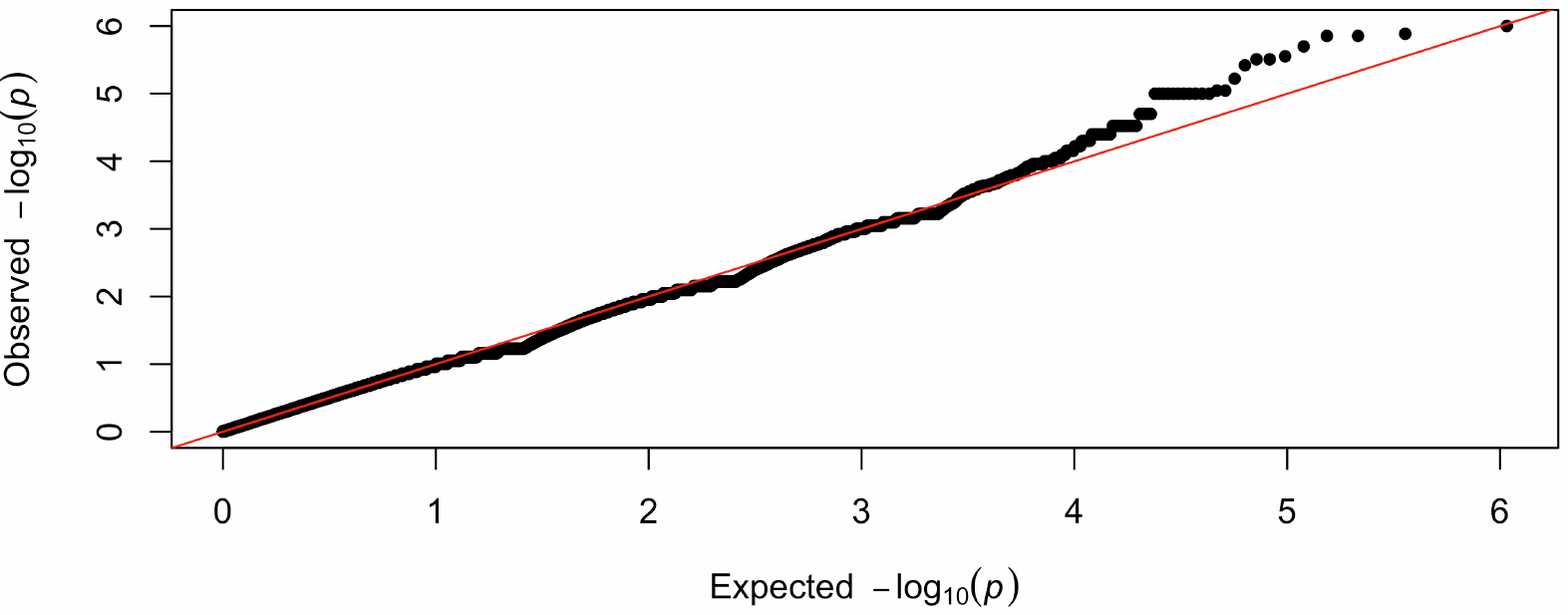}
\caption{QQ-plot of p-values of the MTAF method testing the association between SNPs and multiple traits of substance dependence.
}%
\label{fig:qq}
\end{figure}

\begin{figure}[htbp!]
\centering
\includegraphics[width=\linewidth]{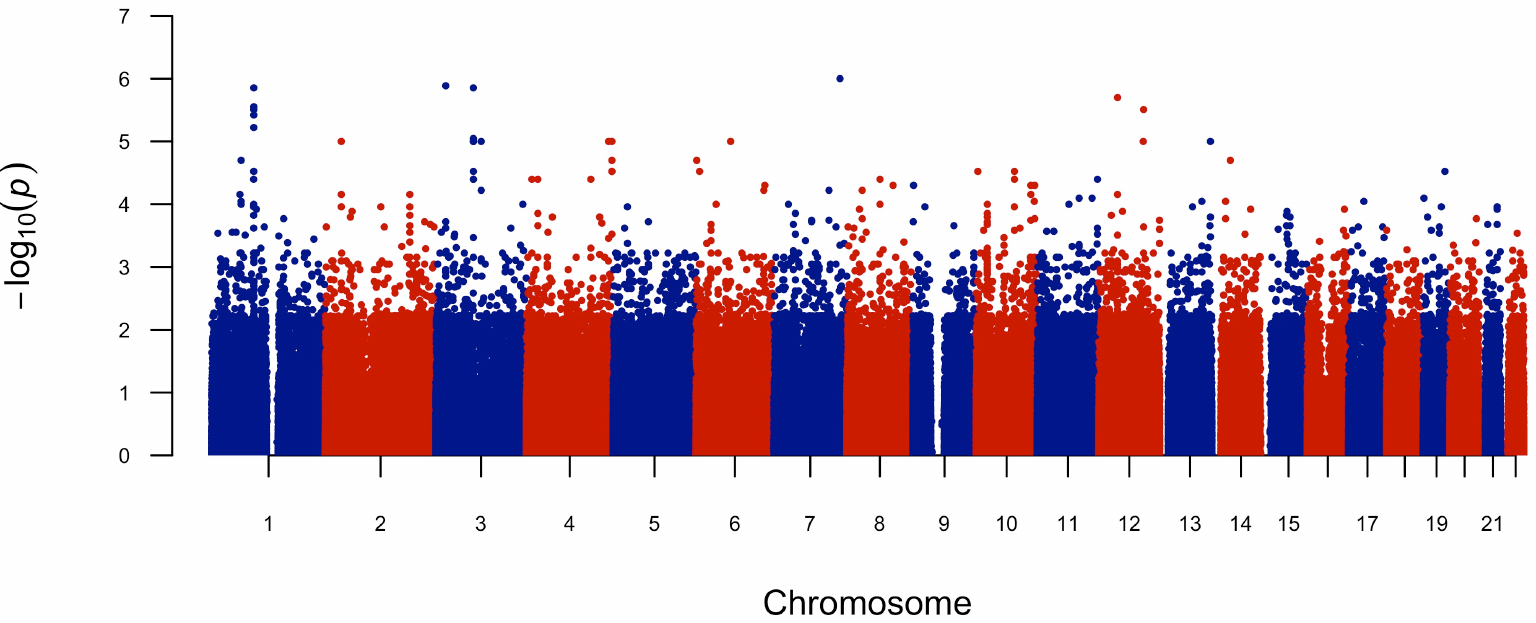}
\caption{Manhattan plot of the MTAF method testing the association between SNPs and multiple traits of substance dependence at a significance level of $5 \times 10^{-6}$
}%
\label{fig:man}
\end{figure}

We inspected the Pearson's correlation among six continuous traits and Table \ref{tab:sage_corr} shows that all six traits are positively correlated. Then we applied the MTAF method to detect the associations between the SNPs and the 18 traits. To get accurate p-values at extremely small significance level (usually lower than $10^{-6}$) given limited computational resource, we performed the tests by adaptively increasing the number of permutations. We first set the number of permutation to be $B$ and filtered out insignificant SNPs with p-values greater than $5/B$ and updated the number of permutation to $10 \times B$. We started with $B = 100$ and repeat the above process until $B=10^7$. By doing this, we managed to avoid permuting $10^7$ times for most of the SNPs, and saved computation resource only for the most significant SNPs. Figure \ref{fig:qq} shows the QQ-plot of the $-\log_{10}(\textrm{p-values})$ of all the SNPs based on the MTAF method. As expected, the majority of the SNPs have p-values that are uniformly distributed, while only a small proportion of the SNPs has strong association with the phenotypes. Please be noted that the inflections observed in the QQ-plot are completely normal due to the adaptive permutation procedure. Figure \ref{fig:man} shows the p-values for all SNPs across 22 chromosomes in a Manhattan plot.

It was known that the SAGE dataset had no SNP achieving $5 \times 10^{-8}$ significance level. Instead, we set the significance level to $5 \times 10^{-6}$ and \citet{yang2016efficient} used the same level when they analyzed the SAGE dataset. At the significance level $5 \times 10^{-6}$, we identified 11 significant SNPs belonging to six genes as shown in Table \ref{tab:sage_SNP}.  Most of these genes are related to the biological functions of nerve and brain, because addictions are considered to be related to mental health. Among these genes, EVI5 is a risk gene of multiple sclerosis (MS), a disease which causes damage to the nervous system \citep{hoppenbrouwers2008evi5, beecham2013analysis}. \citet{gouveia2019genetics} found that ZNF385D is associated with cognitive function, and it was also reported to be associated with language impairment in previous literature \citep{gialluisi2019genome}. TPK1 produces thiamine pyrophosphate and its mutation can cause neurological disorder \citep{banka2014expanding}. According to \citet{davies2018study}, TPK1 is also associated with reaction time. Alzheimer's disease (AD) is associated with LINC02008 which affects the age at onset and MIR4495 which affects memory performance. At last, CNTN1 is associated with AD, Parkinson’s disease (PD), and schizophrenia (SCZ) \citep{vacic2014genome,beecham2014genome,richards2020relationship}.


\begin{table}[htbp]
\caption{\bf significant SNPs at $5 \times 10^{-6}$ level }  
\centering 
\scalebox{0.6}{
\begin{tabular}{c c c c c c} 
\hline 
rsid  & chromosone & position & p-value & gene &  diseases/traits associated

\\ [0.5ex]

\hline 

rs1556562 &1 & 92568466 &$4.0 \times 10^{-6}$ &EVI5  & \\ [0.5ex]
rs1408916 &1 & 92527070 &$3.1 \times 10^{-6}$ &EVI5 & MS \citep{hoppenbrouwers2008evi5, beecham2013analysis} \\ [0.5ex]
rs4847377 &1 & 92557593 &$2.8 \times 10^{-6}$ &EVI5 & \\ [0.5ex]
rs4970712 &1 & 92527990 &$3.8 \times 10^{-6}$ & EVI5 & \\ [0.5ex]
rs9310661 &3 & 21855648 &$1.3 \times 10^{-6}$ & ZNF385D & cognitive function and language impairment \citep{gouveia2019genetics, gialluisi2019genome}\\ [0.5ex]
rs7614064 &3 & 82239346 &$1.4 \times 10^{-6}$ &LINC02008 \\ [0.5ex]
rs7645576 &3 & 82274160 &$3.0 \times 10^{-6}$ &LINC02008 & AD \citep{davies2018study} \\ [0.5ex]
rs9852219 &3 & 82327624 &$5.0 \times 10^{-6}$ &LINC02008 \\ [0.5ex]
rs10224675 &7 & 144595669 &$3.4 \times 10^{-6}$ &TPK1 & thiamine pyrophosphate production and reaction time \citep{banka2014expanding, davies2018study} \\ [0.5ex]
rs11178982 &12 &40907530 &$2.0 \times 10^{-6}$  &CNTN1 & AD, PD, and SCZ \citep{vacic2014genome,beecham2014genome,richards2020relationship} \\ [0.5ex]
rs2020139 &12 &97953656 &$3.1 \times 10^{-6}$ &MIR4495 & AD and memory performance \citep{davies2018study}\\ [0.5ex]

\hline 

\end{tabular}
}
\label{tab:sage_SNP}
\end{table}

\section*{Discussion}
\label{sec:discussion}

Although single-trait analysis methods have been widely used in multiple-trait studies, these methods may be insufficient when traits are truly correlated with each other. Hence, the multiple-trait association tests can increase statistical power by incorporating the information among traits. In this paper, we propose the MTAF method for multiple traits, by adaptively combining p-values from the single-trait analyses.

The MTAF method is very versatile for multiple-trait association testing. First, because the MTAF method only requires the p-values as inputs, it can process both continuous traits and binary traits simultaneously whenever p-values are provided. Second, we apply PCA on continuous traits to uncover hidden components underlying traits, which greatly improves the performance of MTAF when signals are weak and dense. Third, the MTAF method combines the one-sided p-values instead of two-sided p-values, which increases power given the effects of traits are in the same direction. When some of the traits are in the opposite direction of other traits, we can flip these traits such that all or most of the traits are positively correlated.  At last, we paid special attention to the permutation test with covariates. By permuting the residuals of genotypes regressed on the covariates, we managed to control type I error while adjusting for confounders.

Relying on the permutation procedure to get the empirical p-values can be time-consuming when the validity of tiny p-values is required. Since in GWASs most SNPs have no significant effect on the traits, permuting the same number of times on each SNP is unnecessary and can be a waste of computing time and resources. A more efficient way is to permute the data iteratively on each SNP with the expectation that insignificant SNPs will be excluded in less permutation time. As a result, most SNPs get removed after the first few rounds and only significant SNPs require a large number of permutation time. We show the reduction in time complexity by the following example. Starting with $B=100$, the chance that a SNP gets removed in the first round is $0.05$. If the SNP remains, we would start the second round $B=1000$ and the chance that the SNP gets removed in the second round would be $0.045$ which is the difference between the chance of remaining in the first round $0.05$ and the chance of advancing to the third round $0.005$. Following the procedure, if we stop at $B=10^7$, the expected permutation times for a SNP would be $0.95 \cdot 100 + 0.045 \cdot 1000 + \ldots + 4.5 \cdot 10^{-6} \cdot 10^7 \approx 45 \cdot \log_{10}(10^7)$, which is logarithmic time $\log_{10}(B)$. On the contrary, if we fix the permutation times at $10^7$, the expected permutation times would be linear time $B$. Therefore, we reduce the time complexity from linear time to logarithmic time.

A SNP may not always influence the traits directly. Instead, it may indirectly affect correlated traits through an unobserved component, in which case, uncovering the hidden component can enhance the statistical power given correlated traits. In the MTAF method, we use PCA to uncover this potentially hidden component. By detecting the associations between the SNP and hidden components, we may increase the statistical power. This idea was supported by our simulation study in which PCA largely improved the statistical power under dense scenarios. Thus, depending on the correlation structure, PCA may increase the power of testing when traits are correlated. 

Despite the advantages of the MTAF method, several limitations can be addressed in future work. First, the MTAF method can only analyze the single variant at this moment, the set-based analysis is common in GWASs though. \citet{cai2020adaptive} proposed a set-based AF method and the MTAF method might be extended to the set-based analysis or pathway analysis in the future. Second, the p-values are adaptively combined in the MTAF method and we do not aim at selecting the most related traits for the identified SNPs. Thus, in our future work, we can develop a method to provide a list of traits that are most related for each identified SNP. At last, because we need to permute the residuals of genotypes to control for type I error while adjusting for confounders or covariates, the MTAF method requires the individual-level genotype data to perform the test. However, the individual-level data are not always available or often require special permissions since they are considered as identifiable data.  In the future work, we would explore whether the MTAF method can be extended to use the GWAS summary statistics without requiring the individual-level data.

\section*{Acknowledgments}

Funding support for the Study of Addiction: Genetics and Environment (SAGE) was provided through the NIH Genes, Environment and Health Initiative [GEI] (U01 HG004422). SAGE is one of the genome-wide association studies funded as part of the Gene Environment Association Studies (GENEVA) under GEI. Assistance with phenotype harmonization and genotype cleaning, as well as with general study coordination, was provided by the GENEVA Coordinating Center (U01 HG004446). 
Assistance with data cleaning was provided by the National Center for Biotechnology Information. Support for collection of datasets and samples was provided by the Collaborative Study on the Genetics of Alcoholism (COGA; U10 AA008401), the Collaborative Genetic Study of Nicotine Dependence (COGEND; P01 CA089392), and the Family Study of Cocaine Dependence (FSCD; R01 DA013423). Funding support for genotyping, which was performed at the Johns Hopkins University Center for Inherited Disease Research, was provided by the NIH GEI (U01HG004438), the National Institute on Alcohol Abuse and Alcoholism, the National Institute on Drug Abuse, and the NIH contract "High throughput genotyping for studying the genetic contributions to human disease" (HHSN268200782096C). The datasets used for the analyses described in this manuscript were obtained from dbGaP at \url{http://www.ncbi.nlm.nih.gov/projects/gap/cgi-bin/study.cgi?study_id=phs000092.v1.p1} through dbGaP accession number phs000092.v1.p.

\bibliographystyle{plainnat}


\bibliography{mtaf}


\end{document}